\documentclass[aps,twocolumn,prl,preprintnumbers,showpacs]{revtex4}
\usepackage{amsmath}
\usepackage{graphicx} 

\renewcommand\r{\mathbf{x}}
\newcommand\+{\dagger}
\renewcommand\d{\partial}
\newcommand{\gsim}{\hspace*{0.2em}\raisebox{0.5ex}{$>$}
     \hspace{-0.8em}\raisebox{-0.3em}{$\sim$}\hspace*{0.2em}}

\begin{document}

\preprint{INT-PUB 04-11}

\title{Universal Properties of Two-Dimensional Boson Droplets}
\author{H.-W.~Hammer}
\author{D.\,T.~Son}
\affiliation{Institute for Nuclear Theory,
University of Washington, Seattle, Washington 98195-1550}

\begin{abstract}
We consider a system of $N$ nonrelativistic bosons in two dimensions,
interacting weakly via a short-range attractive potential.  We show that
for $N$ large, but below some critical value, the properties of the
$N$-boson bound state are universal.  In particular, the ratio of the binding
energies of $(N+1)$- and $N$-boson systems, $B_{N+1}/B_N$, approaches
a finite limit, approximately 8.567, at large $N$. We also confirm
previous results that the three-body system has exactly two bound
states.  We find for the ground state $B_3^{(0)} = 16.522688(1) B_2$
and for the excited state $B_3^{(1)} = 1.2704091(1) B_2$.
\end{abstract}
\date{October 2004}

\maketitle

The recent experimental progress with ultracold atomic gases has revived the
interest in weakly coupled quantum liquids.  The ability to control
the parameters of the systems make trapped atomic gases ideal laboratories
where theoretical ideas can be checked versus experiment.  One of the
fundamental parameters that can be varied in experiments is the
dimensionality of space.  Both one- and two-dimensional Bose-Einstein
condensates (BEC's) of sodium atoms have been studied in atom 
traps~\cite{Goerlitz}. A one-dimensional condensate of $^7$Li atoms
immersed in a Fermi sea of $^6$Li atoms was observed in \cite{Schreck}.
In Ref.~\cite{Rychtarik}, a two-dimensional BEC of cesium atoms was
realized in a gravito-optical surface trap.
A two-dimensional boson system has also been realized in hydrogen
adsorbed on a helium surface~\cite{Safonov}.

In this paper, we revisit the problem of weakly interacting bosons
in two spatial dimensions (2D).  While most previous theoretical studies
were concerned with a Bose gas with repulsive 
interactions~\cite{Popov,FisherHohenberg}, we focus on attractive
interactions. In particular, we consider a self-bound droplet of $N(\gg1)$
bosons interacting weakly via an {\em attractive}, short-ranged 
pair potential.  Our goal is to exhibit universal properties pertaining 
to large finite systems, which are not in the thermodynamic limit.

We shall show that the system possesses surprising universal
properties.  Namely, if one denotes the size of the $N$-body droplet
as $R_N$, then at large $N$ and in the limit of zero range of the
interaction potential:
\begin{equation}
  \label{RNratio}
  {R_{N+1}}/{R_N} \approx 0.3417,\qquad N\gg 1\,.
\end{equation}
The size of the bound state decreases exponentially with $N$: adding a
boson into an existing $N$-boson droplet reduces the size of the
droplet by almost a factor of three.  Correspondingly, the binding energy of
$N$ bosons $B_N$ increases exponentially with $N$:
\begin{equation}
  \label{BNratio}
  {B_{N+1}}/{B_N} \approx 8.567, \qquad N\gg 1\,.
\end{equation}
This implies that the energy required to remove
one particle from a $N$-body bound state (the analog of the
nucleon separation energy for nuclei) is about 88\% of the
total binding energy.  This is in contrast to most other physical
systems, where separating one particle costs much less energy than the
total binding energy, provided the number of particles in the bound
state is large.

To derive results independent of the details
of the short-distance dynamics such as the ones quoted above,
the $N$-body bound states 
need to be sufficiently shallow and hence have a size
$R_N$ large compared to all other length scales in the problem.
A similar reasoning has
been used in 3D with much success~\cite{relax}.
The breakdown of universality is determined by the next largest length 
scale in the problem, the natural low-energy length scale $\ell$.
Depending on the physical system,
$\ell$ can be the van der Waals length $l_{vdW}$, the range of the 
potential $r_0$ or some other scale. For realistic systems,
Eqs.~(\ref{RNratio}, \ref{BNratio}) 
are valid for large $N$, but below a critical value,
\begin{equation}
  1 \ll N \ll N_{\rm crit} \approx 0.931 \ln({R_2}/{\ell})
  + {\cal O}(1)\,.
\label{eq:break}
\end{equation}
At $N=N_{\rm crit}$ the size of the droplet is comparable to $\ell$
and universality is lost.  If there is a large separation between $R_2$ 
and $\ell$, then $N_{\rm crit}$ is much larger than one and 
the condition (\ref{eq:break}) can be satisfied.
\vspace{6pt}

{\em Asymptotic freedom.}---Our analysis relies strongly on the
property of asymptotic freedom of 2D bosons with attractive
interaction, so we will briefly review this property here.  The system
under consideration is described by the nonrelativistic Hamiltonian
\begin{equation}\label{H}
  H = \int\!d^2\r\, \left( \frac{\hbar^2}{2m}|\nabla\psi|^2 -
      \frac g2 (\psi^\dagger\psi)^2\right).
\end{equation}
The bosons interact via an attractive, short-ranged pair potential
$-g\delta^2(\r)$, with $g>0$.  This choice can be made because at 
low energies, the true potential can not be distinguished from a
$\delta$-function potential. For convenience, we will use the unit system
$\hbar=m=1$; the factors of $\hbar$ and $m$ can be restored
from dimensional analysis. In this unit system, $g$ is
dimensionless, and we assume $g\ll 1$.

In 2D, any attractive potential has at least one bound state.  For the
potential $-g\delta^2(\r)$ with small $g$, there is exactly one bound
state with an exponentially small binding energy,
\begin{equation}\label{B2}
  B_2 \sim \Lambda^2\exp\left( - {4\pi}/g\right),
\end{equation}
where $\Lambda$ is the ultraviolet momentum cutoff (which is the
inverse of the range of the potential).  Equation~(\ref{B2}) can be
obtained by directly solving the Schr\"odinger equation.  However,
this method is not practical for a system with more than a few
particles.

Asymptotic freedom provides an alternative way to understand
Eq.~(\ref{B2}).  In 2D nonrelativistic theory, the four-boson
interaction $g(\psi^\+\psi)^2$ is marginal.  The coupling runs
logarithmically with the length scale $R$, and the running can be
found by performing the standard renormalization group (RG) procedure.
The RG equation reads
\begin{equation}
  \frac{\d g(R)}{\d \ln R} = \frac{g^2(R)}{2\pi}\,.
  \label{RGeq}
\end{equation}
Depending on the sign of $g$, we find two different behaviors.  For
repulsive interactions with $g<0$, the coupling
becomes weaker in the infrared.  For $g>0$, the coupling grows in the
infrared, in a manner similar to the QCD
coupling~\cite{GrossWilczek,Politzer}.  The dependence of the coupling
on the length scale $R$ is given by
\begin{equation}\label{running}
  g(R) = \left[\frac{1}{g} - \frac{1}{4\pi}\ln 
        (\Lambda^2 R^2)\right]^{-1}\,,
\end{equation}
so the coupling becomes large when $R$ is comparable to the size of
the two body bound state $B_2^{-1/2}$.  This is in essence the
phenomenon of dimensional transmutation: a dynamical scale is
generated by the coupling constant and the cutoff scale.

It is natural, then, that $B_2$ is the only physical energy scale in
the problem: the binding energy of three-particle, four-particle,
etc. bound states are proportional to $B_2$.  However, the
$N$-particle binding energy $B_N$ can be very different from $B_2$ if
$N$ is parametrically large.  We shall now argue that $B_N$ increases
exponentially with $N$.

\vspace{2pt}

{\em Stabilizing the size of the droplet.}---We first try to use the
variational method to estimate the size of the bound state.  For a
cluster of a large number of bosons, one can expect that classical
field theory is applicable.  We thus have to minimize the
energy~(\ref{H}) with respect to all field configurations $\psi(\r)$
satisfying the constraint
\begin{equation}
  \label{Ncons}
  N = \int\!d^2\r\, \psi^\+\psi\,.
\end{equation}
It is instructive to first minimize the energy with respect to the
size of the bound state, and afterwards over all shapes of the wave
function.  We use the following trial wave function:
\begin{equation}\label{trialpsi}
  \psi(\r) = \frac{\sqrt N}{R\sqrt{2\pi C}} f\left(\frac rR\right),
\end{equation}
where $r\equiv |{\bf x}|$ and $f(r/R)$ is a function that describes
the shape of the wave function.  We assume that $f(r/R)$ is nonzero
when $r/R\lesssim1$, but becomes small when $r/R\gg1$.  Thus $R$ is the
size of the droplet.  To satisfy the particle number constraint
(\ref{Ncons}), we should take
\begin{equation}\label{C}
  C=\int\!d\rho\,\rho f^2(\rho)\,.
\end{equation}

The total energy is then obtained from Eq.~(\ref{H}) as 
\begin{equation}\label{Ekinpot}
  E(R) = \frac A{2C} \frac N{R^2} - \frac B{4\pi C^2}\frac{gN^2}{R^2}\,,
\end{equation}
where $A$ and $B$ depend on the shape of the wave function,
\begin{equation}\label{AB}
  A = \int\!d\rho\,\rho [f'(\rho)]^2,\qquad
  B = \int\!d\rho\,\rho f^4(\rho)\,.
\end{equation}

As one can see from Eq.~(\ref{Ekinpot}), in 2D both the kinetic and
potential energies scale as $R^{-2}$.  This seems to prohibit a stable
bound state: for $N<2\pi AC/(Bg)$ the system expands to infinite size,
while in the opposite regime, $N>2\pi AC/(Bg)$, it shrinks to zero
size.

However, the above estimate is too crude because it fails to
account for the logarithmic running of the coupling $g$.  We
will therefore replace $g$ in Eq.~(\ref{Ekinpot}) by the coupling at the
length scale $R$: $g\to g(R)$.  This procedure goes beyond the naive
mean-field treatment and captures all leading logarithms (by using the
RG), but it does not take into account all $1/N$ corrections.  We
shall see that this is sufficient for finding the parametric
dependence of $B_N$ on $N$.

Once $g$ is replaced by $g(R)$ in Eq.~(\ref{Ekinpot}), the energy
$E(R)$ has a minimum at a finite $R$.  Indeed, in the limit $R\to0$
the coupling becomes weak, $g\to0$, and $E(R)$ is dominated by the
kinetic energy, which tries to make the system larger.  In the
opposite limit $R\to\infty$ the coupling constant becomes strong, and
$E(R)$ is dominated by the negative potential energy, which favors
smaller $R$.

To find the optimal $R$, we differentiate the energy~(\ref{H}), with
$g$ replaced by $g(R)$, with respect to $R$. We find
\begin{equation}
   AC - \frac{N g(R)}{2\pi}B + \frac{N g^2(R)}{8\pi^2} B = 0\,,
\end{equation}
where we have used Eq.~(\ref{RGeq}).  The solution is
\begin{equation}\label{gR}
  g(R) = \frac{2\pi A C}{NB} + {\cal O}( N^{-2})\,.
\end{equation}
The coupling is ${\cal O}(N^{-1})$, which implies the weak-coupling
regime at large $N$.  The ${\cal O}(N^{-2})$ correction is beyond the
scope of the classical approximation.  Using the RG running of the
coupling constant~(\ref{running}), we find the optimal size of the
droplet,
\begin{equation}\label{R_N}
  R_N = C_R R_2 \exp\left(-\frac{B}{AC}N\right),
\end{equation}
where $C_R$ is some numerical constant of order 1, which cannot be
found at the current level of approximation due to the ${\cal
O}(N^{-2})$ uncertainty in Eq.~(\ref{gR}).  However, we can already
see that the size of the droplet decreases exponentially as a function
of the number of particles.

From Eq.~(\ref{gR}) we see that the kinetic and potential energies
cancel each other to leading order in $1/N$.  For this reason, we can
only estimate the energy to be
\begin{equation}
  B_N = \frac {C_E} {R_N^2} = \frac{C_E}{C_R^2} B_2
   \exp\left(\frac{2B}{AC}N\right)
\label{B_N}
\end{equation}
(barring the possibility that there is a cancellation in the
next-to-leading order in $1/N$), but cannot compute the overall
constant $C_E$.

\vspace{2pt}

{\em The shape of the droplet}.---We now minimize the energy with
respect to the shape of the wave function $f(r/R)$.  Due to the
exponential behavior of the energy as a function of $N$, the optimal
shape is the one which maximizes the ratio $B/(AC)$, where $A$, $B$,
and $C$ are defined in Eqs.~(\ref{C}, \ref{AB}).  This ratio is
truly characteristic of the shape of the wave function---it is
unchanged under the rescaling $f(\rho)\to\lambda_1 f(\lambda_2\rho)$.
The optimal shape of the wave function is therefore ambiguous up to
this trivial rescaling.

Taking the variation of $B/(AC)$ over the $f(\rho)$, we find that
$f(\rho)$ satisfies the equation
\begin{equation}\label{F-eq}
  f''(\rho) + \frac{f'(\rho)}\rho - f(\rho) + f^3(\rho) = 0\,,
\end{equation}
where we have performed the rescaling
\begin{equation}
  f(\rho) \to \sqrt{\frac B{2C}} f\left(\sqrt{\frac AC} \rho\right)\,.
\end{equation}
The boundary condition on $f(\rho)$ is $f'(0)=f(\infty)=0$.  
The solution can be found numerically by using,
e.g., the shooting method.
\begin{figure}[tb]
\centerline{\includegraphics[width=7cm,angle=0]{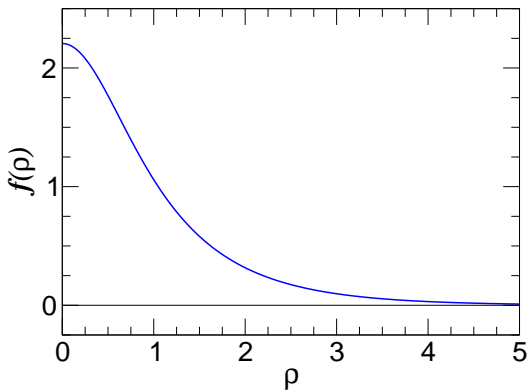}}
\caption{\label{fig0}Numerical solution of Eq.~(\ref{F-eq})
for the  boundary condition $f'(0)=f(\infty)=0$  
obtained using the shooting method.}
\end{figure}
The solution, shown in Fig.~\ref{fig0}, has a characteristic bell
shape with $f(0)\approx2.206$.  For the shape given by the function
$f(\rho)$ solving Eq.~(\ref{F-eq}), $A=\frac12B=C\approx1.862$,
therefore $B/(AC)\approx 1.074$.  Equation~(\ref{B_N}) now can be
written as
\begin{equation}
  B_N = c_1 B_2 c^{N-2}\,,
\end{equation}
where $c\approx8.567$, but $c_1$ is still unknown.
Equations~(\ref{RNratio}) and (\ref{BNratio}) are also recovered.

Equation~(\ref{F-eq}) resembles the Hartree equation for the
single-particle wave function.  The above results can also be obtained
in the Hartree approach, provided the running
coupling is used instead of the bare one.

\vspace{2pt}

{\em Three-body bound state}.---We next describe our computation of
the binding energies of the three-body system, which can be calculated
exactly. For this purpose, we use an effective field theory and work 
in the Lagrangian formalism. It is convenient to introduce an
auxiliary field $d\equiv\psi^2$ with the quantum numbers of 
two bosons (sometimes called the \lq\lq dimeron'')
\cite{Kaplan:1996nv,3bdy:3d}. In terms 
of $d$ and $\psi$, the Lagrangian density corresponding to Eq.~(\ref{H})
reads
\begin{equation}
{\cal L} = \psi^\dagger \left( i\partial_t +\frac{\nabla^2}{2}\right)\psi 
      -\frac{g}{2}\,d^\dagger d+\frac{g}{2}\left( 
        d^\dagger \psi^2+{\psi^\dagger}^2 d\right).
\end{equation}
The boson propagator takes the usual nonrelativistic form 
$i/(p_0-{\bf p}^2/2+i\epsilon)$. It is not renormalized by interactions
since all tadpole diagrams vanish in this theory.
The bare dimeron propagator is simply a constant $-2i/g$. 
In the presence of interactions, it gets dressed by boson bubbles 
to all orders, leading to the full propagator:
\begin{equation}
i\Delta(p_0,{\bf p})=-i\frac{8 \pi}{g^2}\ln\left[
  \frac{{\bf p}^2/4-p_0-i\epsilon}{B_2}\right]^{-1},
\end{equation}
where the bare coupling constant $g$ will drop out of all observables
in the end. The Feynman rule for the $d\psi\psi$ vertex coupling the 
dimeron to two bosons is $ig$.

\begin{figure}[tb]
\vspace*{0.3cm}
\centerline{\includegraphics[width=7cm,angle=0]{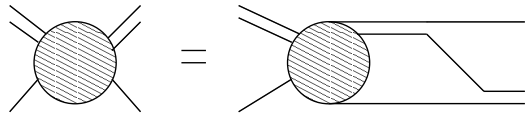}}
\caption{\label{fig1}The integral equation for the 
three-body amplitude. The single (double) line indicates the
boson (full dimeron) propagators, respectively.
}
\end{figure}

The three-body binding energies are determined by the 
homogeneous integral equation for the three-body bound state amplitude
depicted in Fig.~\ref{fig1}.
The single (double) line indicates the boson (full dimeron) propagators, 
respectively, while the blob is the bound state amplitude. It
depends on the total energy $E$ and the relative momentum of the 
boson and the dimeron. 
The three-body binding energies are given by those (negative)
values of the total energy $E=-B_3$, for which the homogeneous integral 
equation shown in Fig.~\ref{fig1} has a nontrivial solution.
The derivation of the integral equation using the Feynman rules given
above proceeds as in the three-dimensional case \cite{3bdy:3d}. There
are only bound states if the dimeron and the third boson are in 
a relative S-wave. The formation of bound states in the higher partial waves
is prevented by the angular momentum barrier.
Projecting onto the S-wave, we obtain the equation for the 
bound state amplitude $F(p)$:
\begin{equation}
\label{3bdyeq}
F(p)=\int_0^\infty \frac{4q\, dq\; F(q)\;\;
  \ln\left[(3q^2/4+B_3)/B_2\right]^{-1}}
  {\sqrt{(p^2+q^2+B_3)^2-p^2 q^2}}\,.
\end{equation}
The three-body binding energies can be obtained numerically
to high precision by discretizing Eq.~(\ref{3bdyeq}).

We find exactly two three-body bound states: the ground state with
binding energy $B_3^{(0)} = 16.522688(1) B_2$ and one excited state
with $B_3^{(1)} = 1.2704091(1) B_2$. (The numbers in parentheses give
the error in the last digit.)  
The three-body binding energies for a
zero-range potential in 2D have previously been calculated by 
Bruch and Tjon \cite{Bruch} and Nielsen et al.~\cite{Nielsen}.
Our results are consistent with the previous calculations
but more precise. Platter et al. have recently calculated 
the four-body binding energies for a zero-range potential in 2D 
\cite{Platter}. They found  exactly two bound states: the ground 
state with $B_4^{(0)}=197.3(1)B_2$ and one excited state with 
$B_4^{(1)}=25.5(1)B_2$.

The results for the ground state energies $B_3^{(0)}$ and $B_4^{(0)}$
are what should be compared with the asymptotic formula~(\ref{BNratio}). 
The ratio $B_3^{(0)}/B_2\approx 16.5$ is almost twice as large as the 
asymptotic value~(\ref{BNratio}), while the ratio $B_3^{(0)}/B_4^{(0)}
\approx 11.9$ is considerably closer.
Such deviations are expected for the small 
values of $N$ we are dealing with. Note, however, that the ratio of
the root mean square radii of the two- and three-body wave functions
is $0.306$~\cite{Nielsen}, close to the asymptotic
value~(\ref{RNratio}).

\vspace{2pt}

{\em Conclusion}.---We have evaluated parametrically the bound state
energy of of $N$ weakly attracting nonrelativistic bosons.  Our
results are obtained by minimizing the mean-field energy functional
with a scale-dependent coupling.  While our approximation is good
enough to establish the exponential behavior of the binding energy on
the number of particles, it is not sufficiently accurate for
evaluating the overall coefficient in front of the exponent.  It would
be valuable to develop a technique capable of doing so.

We also have computed the binding energy of a system of three bosons.  We
have confirmed the previous finding that there are two bound
states~\cite{Bruch,Nielsen} and improved the precision of the universal
values for their binding energies. One also would like to 
directly compute the ground state energy of the $N$-body system for $N>4$
and compare the ratio $B_N^{(0)}/B_{N-1}^{(0)}$ with the asymptotic value.  
It would be also interesting to know the number of excited states of
the $N$-body bound system at large $N$.

The many-particle bound state studied here is analogous in to the
nontopological soliton, or Q-ball~\cite{Friedberg:me,Coleman} in
relativistic quantum field theory.  Nonrelativistic bosons in 2D
provide an interesting example where the size of the nontopological
soliton is stabilized by a pure quantum effect (the running of the
coupling).

An important question to explore is whether the result of this work
can be extended to three-dimensional boson systems with large
scattering length~\cite{Bulgac}.  In contrast to the 2D case, this
system displays the Efimov effect~\cite{Efimov71}, and the three-body
bound state energy $B_3$ is an independent parameter. On the other
hand, the four-body bound state energy can be expressed via $B_2$ and
$B_3$~\cite{Platter:2004qn}.  One would like to know if the binding
energy of $N$ bosons can be expressed in terms of $B_2$ and $B_3$
alone and find the large $N$ behavior of $B_N$ and the wave function.

Finally, one should investigate the realizablity of self-bound 
2D boson systems with weak interactions in experiments. According to
the analysis of Ref.~\cite{Platter}, the 
$1/N$ corrections to Eqs.~(\ref{RNratio}, \ref{BNratio}) are small
for $N\gsim 6$. Using (\ref{eq:break}),
this requires $R_2/\ell \gg 600$. We are not aware of
any physical system that satisfies this constraint. However, such
a system could possibly be realized close to a Feshbach resonance
where $R_2$ can be made arbitrarily large. A interesting
theoretical question is the dynamics of the droplet formation
\cite{Josserand}.

\vspace{2pt}

We thank G.~Bertsch, E.~Braaten, D.\,B.~Kaplan, 
and Yu.\,V.~Kovchegov for discussions. This re\-search was 
supported by DOE grant DE-FG02-00ER41132.
D.T.S. was supported, in part, by the Alfred P.~Sloan Foundation.

\end{document}